\documentclass[a4paper]{revtex4}
\usepackage{graphicx}
\usepackage{fancyhdr}
\usepackage{amsmath}
\usepackage{graphicx,color}
\usepackage{amssymb}
\newcommand{\VEV}[1]{\langle #1 \rangle}

\newcommand{\diag}{\mathrm{diag}}

\def\fsl#1{\setbox0=\hbox{$#1$}                 
   \dimen0=\wd0                                 
   \setbox1=\hbox{/} \dimen1=\wd1               
   \ifdim\dimen0>\dimen1                        
      \rlap{\hbox to \dimen0{\hfil/\hfil}}      
      #1                                        
   \else                                        
      \rlap{\hbox to \dimen1{\hfil$#1$\hfil}}   
      /                                         
   \fi}                                         %

\begin{document}

\title{St\"{u}ckelberg model and Composite $Z'$}

%

\author{Michio Hashimoto}
\affiliation{Chubu University, 1200 Matsumoto-cho, 
   Kasugai-shi,  Aichi, 487-8501, JAPAN}

\begin{abstract}
Based on Ref.~\cite{Hashimoto:2014lfa}, 
we study a composite $Z'$ model which effectively induces 
the St\"{u}ckelberg model in low energy.
It turns out that the mass of the composite $Z'$ boson contains
the St\"{u}ckelberg mass term in sharp contrast to 
the conventional $Z'$ model.
We also find that the masses of the composite scalar and 
the right-handed neutrinos are determined by the infrared fixed points.
If future experiments confirm that the gauge coupling $g$ of $Z'$ is 
sufficiently large, say, $g^2/(4\pi) \gtrsim 0.015$ for the $U(1)_{B-L}$ model,
and also establish the existence of the St\"{u}ckelberg mass term for $Z'$,
it might be evidence of the compositeness of $Z'$.
\end{abstract}

\maketitle

\thispagestyle{fancy}

\section{Introduction}

The standard model (SM) was almost confirmed by the discovery of
the Higgs boson~\cite{Higgs-discover}.
It turns out that the perturbation theory works up to
high energy near the  Planck scale.
Is there a room for strongly interacting theories such as 
(walking) Technicolor, top condensate and 
other models~\cite{Hill:2002ap,MHashimoto:works}?

We here explore a possibility of a composite $Z'$ which 
effectively induces the St\"{u}ckelberg model 
in low energy~\cite{Stueckelberg:1900zz,Ruegg:2003ps}.
If the strong coupling region is around the Planck or the GUT scale,
a big $U(1)$ gauge coupling $g$ is not necessarily needed in low energy.
For the $U(1)_{B-L}$ model, $g^2/(4\pi) \gtrsim 0.015$ is sufficient.
We find that the masses of the extra scalar and 
the right-handed neutrino are controlled by the infrared fixed points. 
In sharp contrast to the conventional $U(1)_{B-L}$ model, 
the $Z'$ mass inevitably has the contribution of the St\"{u}ckelberg mass term
in the composite $Z'$ model.
This extra contribution to the $Z'$ mass might be 
the remnant of the strong dynamics in high energy.

\section{St\"{u}ckelberg model and composite vector boson}
\label{sec2}

Let us start from a model with a Majorana-type scalar four-fermion coupling
and a vector one:
\begin{equation}
  {\cal L} = \bar{\eta} i \fsl{\partial} \eta
  + G_S (\overline{\eta^c} \eta)(\overline{\eta} \eta^c)
  - G_V (\overline{\eta}\gamma^\mu \eta)^2,
  \label{NJL-Maj}
\end{equation}
where $\eta$ is a two-component fermion,
for example, a right-handed neutrino, and
$\eta^c$ is the charge conjugation.
By introducing composite scalar and vector fields,
$\phi \sim \overline{\eta} \eta^c$,
$\phi^\dagger \sim \overline{\eta^c} \eta$,
and $A_\mu \sim \overline{\eta}\gamma^\mu \eta$,
we can rewrite the theory in terms of the system of 
the fermion, and the composite scalar and vector bosons.

In low energy, the composite scalar and vector fields acquire 
the kinetic terms via the bubble diagrams.
Then the induced effective theory in a low energy scale $\mu$ is
\begin{eqnarray}
 {\cal L}_{\rm eff} &=& \bar{\eta} i\fsl{D}\eta
  + Z_\phi |D_\mu \phi|^2 - M_\phi^2 \phi^\dagger \phi 
  - \lambda_\phi (\phi^\dagger \phi)^2 
  - \overline{\eta^c} \eta \phi 
  - \overline{\eta} \eta^c \phi^\dagger 
  - \frac{Z_A}{4} F_{\mu\nu} F^{\mu\nu}
  + \frac{1}{2} f^2 A_\mu^2,
\end{eqnarray}
where $D_\mu \eta = \partial_\mu \eta - i A_\mu \eta$, 
$D_\mu \phi = \partial_\mu \phi + 2i A_\mu \phi$,
and the scalar quartic coupling $\lambda_\phi$ is also induced 
by the bubble diagram.
The wave function renormalization constants are
\begin{equation}
  Z_\phi = \frac{1}{16\pi^2} \log \Lambda^2/\mu^2, \quad
  Z_A = \frac{1}{24\pi^2} \log \Lambda^2/\mu^2 ,
\end{equation}
where we used the proper time regularization.
Introducing $g \equiv Z_A^{-1/2}$ and $y \equiv Z_\phi^{-1/2}$,
and rescaling $A_\mu$ and $\phi$ as $A_\mu \to g A_\mu$ and $\phi \to y\phi$,
respectively,
the effective theory has the canonical kinetic terms, 
\begin{eqnarray}
 {\cal L}_{\rm eff} &=& \bar{\eta} i\fsl{D}\eta
  + |D_\mu \phi|^2 - \tilde{M}_\phi^2 \phi^\dagger \phi 
  - \tilde{\lambda}_\phi (\phi^\dagger \phi)^2 
  - y \overline{\eta^c} \eta \phi 
  - y \overline{\eta} \eta^c \phi^\dagger 
  - \frac{1}{4} F_{\mu\nu} F^{\mu\nu}
  + \frac{1}{2} g^2 f^2 A_\mu^2 \, .
\end{eqnarray}
The field-dependent rotations for
the fermion and scalar variables,
\begin{equation}
  \varphi \equiv e^{i\frac{B(x)}{gf}} \eta, \quad
  \overline{\varphi} \equiv e^{-i\frac{B(x)}{gf}} \overline{\eta},
  \quad
  \chi \equiv e^{-2i\frac{B(x)}{gf}} \phi, \quad
  \chi^\dagger \equiv e^{2i\frac{B(x)}{gf}} \phi^\dagger, 
  \label{Weinberg-rotation}
\end{equation}
and the redefinition of the gauge field 
$\tilde{A}_\mu \equiv A_\mu + \frac{1}{gf} \partial_\mu B$ yield
\begin{eqnarray}
 {\cal L}_{\rm eff} &=&
    \bar{\varphi} (i\fsl{\partial} + g \fsl{\tilde{A}}) \varphi
  + |(\partial_\mu + 2i g \tilde{A}_\mu) \chi|^2
  - \tilde{M}_\chi^2 \chi^\dagger \chi 
  - \lambda (\chi^\dagger \chi)^2 
  - y \overline{\varphi^c} \varphi \chi 
  - y \overline{\varphi} \varphi^c \chi^\dagger \nonumber \\
&&
  - \frac{1}{4} F_{\mu\nu} F^{\mu\nu}
  + \frac{1}{2} g^2 f^2
    \left(\tilde{A}_\mu - \frac{1}{gf} \partial_\mu B\right)^2 \, .
    \label{Leff}
\end{eqnarray}
It is nothing but the St\"{u}ckelberg model 
with the complex scalar field~\cite{Stueckelberg:1900zz,Ruegg:2003ps}.
Since we introduced the redundant field $B(x)$,
we should add a delta function $\delta (\xi_B -1)$ with
$\xi_B \equiv e^{i\frac{B(x)}{gf}}$ in the path integral,
which is connected with the gauge fixing term.
Note that we cannot avoid quadratically fine-tuning
to the mass terms of the composite scalar and vector fields 
in this Nambu-Jona-Lasinio (NJL) picture.

In this way, by introducing the St\"{u}ckelberg scalar field $B(x)$
as in Eq.~(\ref{Weinberg-rotation}),
the original global $U(1)$ symmetry in Eq.~(\ref{NJL-Maj})
is upgraded to the local one in Eq.~(\ref{Leff}).
We also find that
the St\"{u}ckelberg model as a low energy effective theory
corresponds to the composite model in a high energy scale $\Lambda$, 
when we impose the compositeness conditions
in the context of the renormalization group equations 
(RGE's)~\cite{Bardeen:1989ds}, 
\begin{equation}
  \frac{1}{g^2 (\Lambda)} = \frac{1}{y^2 (\Lambda)} = 0, \quad
  \frac{\lambda (\Lambda)}{y^4 (\Lambda)} = 0 \, .
\end{equation}

\section{Composite $Z'$ model}

Let us study the $U(1)_{B-L}$ extension of the SM:
\begin{equation}
  {\cal L} = {\cal L}_{\rm SM} + {\cal L}_{\nu} + {\cal L}_\chi
  + {\cal L}_{Z'} + {\cal L}_{\rm gf},
\end{equation}
where ${\cal L}_{\rm SM}$ represents the SM part, and
\begin{equation}
  {\cal L}_\nu = \sum_{f=1,2,3} \overline{\nu_R^f} i\fsl{D} \nu_R^f
\end{equation}
\begin{eqnarray}
  {\cal L}_\chi &=& |D_\mu \chi|^2 - M_\chi^2 \chi^\dagger \chi
  - \lambda_\chi (\chi^\dagger \chi)^2 - \lambda_{\chi H} |H|^2 |\chi|^2
  - Y_{jk} \overline{\nu_R^{j\,c}}\nu_R^k \chi
  - Y_{jk} \overline{\nu_R^j}\nu_R^{k\,c} \chi^\dagger, 
\end{eqnarray}
\begin{equation}
  {\cal L}_{Z'} = - \frac{1}{4} F_{\mu\nu} F^{\mu\nu}
  + \frac{1}{2} g^2 f^2 \left(A_\mu - \frac{1}{gf} \partial_\mu B\right)^2\,.
  \label{Zp}
\end{equation}
The SM Higgs doublet and the gauge fixing term are denoted by
$H$ and ${\cal L}_{\rm gf}$, respectively.
The $U(1)$ part of the covariant derivative is 
\begin{eqnarray}
  D_\mu &=& \partial_\mu 
  - i Q_Y (g_Y Y_\mu + \tilde{g} A_\mu)
  - i g Q_{B-L} A_{\mu}, 
\end{eqnarray}
where $Q_Y$ and $Q_{B-L}$ represent the hypercharge and the $B-L$ charge, 
respectively.
The $U(1)_Y$ and $U(1)_{B-L}$ gauge couplings are $g_Y$ and $g$, respectively.
Although the gauge mixing coupling $\tilde{g}$ appears in general,
we set $\tilde{g}(\Lambda) = 0$, because there is
no gauge kinetic mixing term at the compositeness scale $\Lambda$.
Noting that the operator $|H|^2 |\chi|^2$ has a higher dimension than six
at the compositeness scale $\Lambda$,
we may neglect the scalar quartic mixing $\lambda_{\chi H}$ at $\Lambda$;
i.e., we also set $\lambda_{\chi H}(\Lambda)=0$. 

In Eq.~(\ref{Zp}), the St\"{u}ckelberg mass term is incorporated from
the beginning unlike the conventional $Z'$ model. 
The existence of this term is essential in our formalism of 
the composite vector field.

The full set of the RGE's for the $U(1)_{B-L}$ model is
shown in Refs.~\cite{Iso:2009ss,Hashimoto:2013hta}.

The key points of the RGE's for the gauge and Yukawa couplings are
\begin{eqnarray}
  \beta_g &\equiv& \mu \frac{\partial }{\partial \mu} g =
  \frac{a}{16 \pi^2}  g^3, \\
  \beta_y &\equiv& \mu \frac{\partial }{\partial \mu} y =
  \frac{y}{16 \pi^2} \bigg[\, b y^2 - c g^2\,\bigg], 
\end{eqnarray}
with $a=12$, $b=10$, and $c=6$,
where we took $Y_{jk} = \diag (y,y,y)$.
The compositeness conditions, $1/g^2(\Lambda)=1/y^2(\Lambda)=0$, yield
\begin{equation}
  \frac{1}{g^2(\mu)} = \frac{a}{8\pi^2} \ln \frac{\Lambda}{\mu}, \quad
  \frac{1}{y^2(\mu)} = \frac{b}{a+c}\,\frac{1}{g^2 (\mu)} ,
  \label{PRFP}
\end{equation}
where $\Lambda$ is the compositeness scale.
Note that the solution (\ref{PRFP}) corresponds to the infrared fixed point.
In fact, we can easily rewrite the RGE's as follows:
 \begin{equation}
  (8\pi^2)\,\mu \frac{\partial }{\partial \mu} \left(\frac{y^2}{g^2}\right)
  = b \, g^2 \cdot \frac{y^2}{g^2}
    \left(\,\dfrac{~y^2~}{g^2} - \frac{a+c}{b}\,\right),
  \label{RGE-PR}
\end{equation}
which is similar to
the Pendleton--Ross type~\cite{Pendleton:1980as}.
Strictly speaking, the asymptotic free theory ($a < 0$)
was studied in Ref.~\cite{Pendleton:1980as}. 
Thus the situation $1/g^2(\Lambda) \to 0$ occurs in low energy
unlike in the asymptotic nonfree theory ($a > 0$).
Owing to the nature of the infrared fixed point, 
even if we relax the compositeness conditions
to the nonvanishing ones, $1/g^2(\Lambda), 1/y^2(\Lambda) \ll 1$, 
the RG flows are not changed so much. 

The RGE for $\lambda_\chi$ is a bit complicated:
\begin{eqnarray}
    \beta_{\lambda_\chi} &\equiv&
    \mu \frac{\partial }{\partial \mu} \lambda_\chi =
    \frac{1}{16 \pi^2} \bigg[\, 20 \lambda_\chi^2
    + \lambda_\chi (24 y^2 - 48 g^2) - 48 y^4 + 96 g^4\,\bigg], 
\end{eqnarray}
where we ignored the numerically irrelevant $\lambda_{\chi H}^2$ term.
Substituting the solutions (\ref{PRFP}) for $g$ and $y$, 
we obtain
\begin{eqnarray}
  (16\pi^2) \mu \frac{\partial }{\partial \mu}
  \left(\frac{\lambda_\chi}{g^2}\right) 
 = 20 g^2 \left(\frac{\lambda_\chi}{g^2} - k_+ \right)
     \left(\frac{\lambda_\chi}{g^2} - k_- \right),
\end{eqnarray}
where $k_+ \equiv \frac{2}{25} \big(9+\sqrt{546}\big) \simeq 2.589$ and 
$k_- \equiv \frac{2}{25} \big(9-\sqrt{546}\big) \simeq -1.149$.
Thus the solution $\lambda_\chi/g^2 = k_+$ is an infrared fixed point.
We can confirm that the analytical expression of 
the solution for $\lambda_\chi$ with the compositeness condition,
$\lambda_\chi(\Lambda)/y^4(\Lambda) = 0$, is actually
\begin{equation}
  \lambda_\chi (\mu) = \frac{2}{25} \big(9+\sqrt{546}\big) \,g^2(\mu) ,
  \label{sol-lam}
\end{equation}
where we assumed positivity of $\lambda_\chi$ in any scale.

The Majorana Yukawa couplings and the quartic coupling of 
the extra composite scalar are proportional to the $U(1)$ gauge coupling
and the coefficients are determined through the infrared fixed points.
As a result, the mass ratio of $\nu_R$ and $\chi$ is controlled by 
the nature of the infrared fixed point.

Let us take the vacuum expectation value (VEV) of $\chi$ 
as $\VEV{v_\chi} = v_\chi/\sqrt{2}$.
Then the square of the masses of $\nu_R$, $\chi$ and $Z'$ are
\begin{equation}
 M_{\nu_R}^2 \simeq 2 y^2 v_\chi^2, \quad
 M_\chi^2 \simeq 2\lambda_\chi v_\chi^2, \quad 
 M_{Z'}^2 \simeq 4 g^2v_\chi^2 + g^2 f^2 \, .  
\end{equation}
We thereby find the mass relation between $\nu_R$ and $\chi$ as
\begin{equation}
  \frac{M_\chi}{M_{\nu_R}} = \frac{\sqrt{\lambda_\chi}}{y} 
  = \frac{\frac{\sqrt{2(9+\sqrt{546})}}{5}}{\frac{3}{\sqrt{5}}}
    \approx 1.2,
\end{equation}
owing to the nature of the infrared fixed points.
In sharp contrast to the conventional approach for $Z'$,
we have the contribution of the St\"{u}ckelberg mass to $M_{Z'}$,
\begin{equation}
  \Delta \equiv \frac{M_{Z'}^2}{g^2} - 4v_\chi^2 = f^2 > 0 \,.
\end{equation}
If the experiments such as LHC and ILC observe $\Delta > 0$
and confirm $g^2/(4\pi) \gtrsim 0.015$,
it implies the compositeness of $Z'$.

\section{Summary and discussions}

We investigated the possibility of the composite $Z'$.
We showed 
that the NJL model effectively induces 
the St\"{u}ckelberg model in low energy via the fermion bubble diagrams.
In terms of the RGE's, this correspondence is realized by
the compositeness conditions.
We also showed that the RG flows are essentially controlled by 
the infrared fixed points.
The nature of the infrared fixed points gives the mass ratio,
$M_\chi/M_{\nu_R} = \sqrt{\lambda_\chi}/y \approx 1.2$.
In the composite $Z'$ model, there are two contributions to
the $Z'$ mass: First is the VEV of $\chi$, which is the conventional one, and 
the second is the St\"{u}ckelberg mass term.
If $g^2/(4\pi) \gtrsim 0.015$ is confirmed and also 
the existence of this extra mass term,
$\Delta \equiv M_{Z'}^2/g^2 - 4v_\chi^2 > 0$, is established 
in future experiments~\cite{Feldman:2006wb}, 
it will be an evidence of the strong dynamics in high energy.

The scenario that the composite $Z'$ boson generated around 
the Planck or the GUT scale survives in low energy, of course,
suffers from the naturalness problem. 
On the other hand, we may consider a scenario that 
the masses of $Z'$, $\chi$ and $\nu_R$ are not 
so far below the compositeness scale $\Lambda$.
In this case, the seesaw mechanism~\cite{seesaw} can work.
We here point out that the SM Higgs potential can be stabilized
by the tree level shift of the Higgs quartic coupling essentially
generated by the $Z'$ loop contribution~\cite{EliasMiro:2012ay}. 
We will study such a scenario elsewhere.

\bigskip 

\end{document}